# The Prototype of Decentralized Multilateral Co-Governing Post-IP Internet Architecture and Its Testing on Operator Networks

Hui Li[1], Jiangxing Wu[2], Kaixuan Xing[1*], Peng Yi[2], Julong Lan[2], Xinsheng Ji[2], Qinrang Liu[2], Shisheng Chen[3], Wei Liang[3], Jinwu Wei[4], Wei Li[4], Fusheng Zhu[5], Kaiyan Tian[6], Jiang Zhu[6], Yiqin Lu[7], Ke Xu[8], Jiaxing Song[8], Yijun Liu[9], Junfeng Ma[10], Rui Xu[1], Jianming Que[1], Weihao Yang[11], Weihao Miu[12], Zefeng Zheng[13], Guohua Wei[1], Jiuhua Qi[1], Yongjie Bai[1], Chonghui Ning[1], Han Wang[1], Xinchun Zhang[1], Xin Yang[1], Jiansen Huang[1], Sai Lv[1], Xinwei Liu[1], Gengxin Li[1]

[1]Shenzhen Graduate School, Peking University, P. R. China
[2]National Digital Switching System Engineering and Technological Research Center, P.R. China
[3]China Telecom Corporation Limited, P.R. China
[4]China United Network Communications Limited, P.R. China
[5]Guangdong Communications & Networks Institute, P.R. China
[6]Kingsoft Cloud Network Technology Co., Ltd. P.R. China
[7]South China University of Technology, P.R. China
[8]Tsinghua University, P.R. China
[9]Guangdong University of Technology, P.R. China
[10]The China Academy of Information and Communications Technology, P.R. China
[11]The Chinese University of Hong Kong, P.R. China
[12]The Hong Kong University of Science and Technology, P.R. China
[13]Macau University of Science and Technology, P.R. China

\* Corresponding author. E-mail: kaixuan@sz.pku.edu.cn

*Abstract*—Internet has become the most important infrastructure of modern society, while the existing IP network is unable to provide high-quality service. the unilateralism IP network is unable to satisfy the Co-managing and Co-governing demands to Cyberspace for most Nations in the world as well. Facing this challenge, we propose a novel Decentralized Multilateral Co-Governing Post-IP Internet architecture. To verify its effectiveness, we develop the prototype on the operator's networks including China Mainland, Hong Kong, and Macao. The experiments and testing results show that this architecture is feasible for co-existing of CCN (Content-Centric Networking) and IP network, and it might become a Chinese Solution to the world.

*Keywords—Multilateral governing; DNS; Multi-modal network; Multi-identifier network; Alliance chain*

## I. INTRODUCTION

IP network is an essential component in modern society. Traditional Internet exposes innate deficiencies as the amount of network information and identifiers grows exponentially with the large-scale deployment and application of big data, mobile internet and the Internet of Things. Restrictions such as the depletion of resources and poor business adaptability hinder traditional IP networks for further development. The existing IP network is unable to satisfy the demand in Co-managing and Co-governing to cyberspace for most nations.

In recent years, there are several architectures proposed which are independent of IP by various countries around the world. In addition, resources and network identifiers cannot be limited to the IP domain name in the future network. The consensus is that the network in the post-IP era should support multi-identifiers including content, ID, IP address, and geospatial location. Multi-modal identifiers and domain name in cyberspace have to be shared by all humankind.

In this paper, while analyzing and considering the problems of existing network architecture on the trend of Internet in high security, high robustness, high efficiency and high availability, we propose a new domain name and identifiers management architecture combining the characteristics of multilateral governing, multi-modal addressing, endogenous security, efficient availability and privacy protection. The prototype system implements the mechanism of binding user content with its private key signature, as well as the multi-identifiers accessed under several transmission scenarios such as IP-CCN-IP, IP-CCN, CCN-IP, CCN-IP-CCN, and CCN-CCN. The prototype system has achieved real-world video stream transmission in heterogeneous transmission channels. The experiments and testing of this prototype on the operator's networks including Mainland China, Hong Kong, and Macao show that this architecture is feasible for co-existing of CCN and IP network.

## II. BACKGROUND

In 2000, ICANN deployed 13 root servers around the world. RFC2535 claimed that the number of root servers could not be further expanded due to the limitation of the number of bytes. Key resources such as root servers, domain names and AS numbers are still controlled by the National Telecommunications and Information Administration, a division of the US Commerce Department. The centralized controlled architecture brings potential threats to global Internet security. In 2002, the DNS root server suffered a large-scale DDoS attack, which seriously affected the global domain name resolution service. In 2014, DNS service in China broke down because of cache pollution. In 2015, nearly all Turkey's (.tr) domains were inaccessible because of the national top-level domain was attacked.

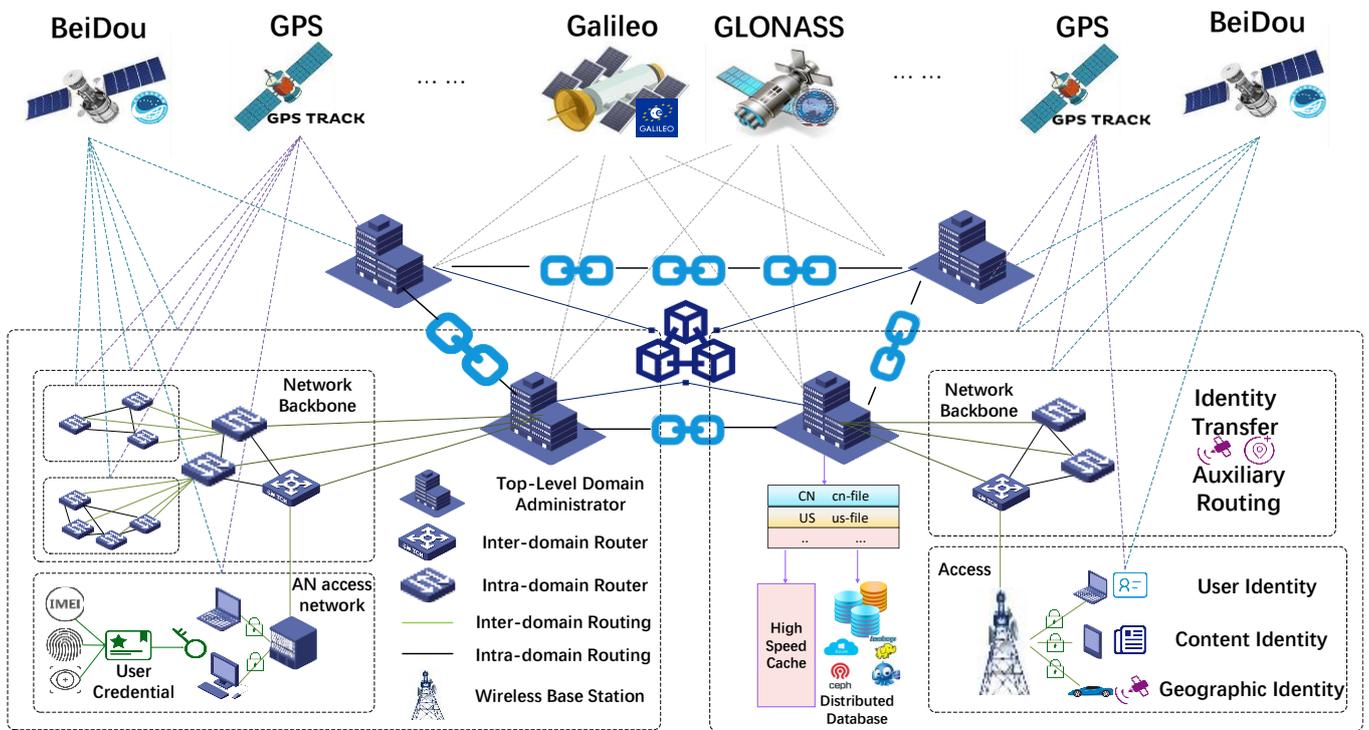

Figure 1. Overall architecture of new domain name system

In August 2010, the national science foundation of the United States announced the FIA Future Internet Architecture project, funding four future network research: Named Data Network, MobilityFirst, NEBULA, XIA (eXpressive Internet Architecture). NDN project aims to develop a new network architecture and establish a content-based Internet architecture. MobilityFirst solves the mobility and reliability problems by treating the mobile nodes as common nodes. Delay Tolerant Network Technology (DTN) is applied to enhance the robustness and usability of the network. With the joint participation of the US national science foundation and top 18 universities and experimental institutions, FIA project stepped into FIA-NP in 2015, planning to establish Internet architecture covering the whole world based on the above network innovation.

In 2012, the European Union funded about 25 million Euros for the research of the future network architecture and service mechanism based on the previous FIRE (Future Internet Research and Experimentation) program. This research tried to make the network more intelligent by exploring self-cognition management mechanism. In the same year, Japan's AKARI as well as the German G-lab program carried out the second phase of the construction of the testbed. However, neither project has been designed for future network architecture.

China has support for the research on basic theories of new-generation networks. In 2007, National Program on Key Basic Research Project of China (973 Project) funded the project of "Basic Research on Testable, Controllable and Manageable IP Network", which mainly conducted research on testability, controllability and manageability of existing IP networks. In 2008, the National 973 Project funded the project "Basic Research on New-Generation Internet Architecture and Protocols". In 2011, 973 Project supported "The Future Internet Architecture and The Mechanism of Service-Oriented" and "Reconfigurable Information Communication Network System Research"[1]. These projects gives direction towards future network research [2][3].

Consider the fact that the previously mentioned projects all still have the problem of centralizing architecture. A decentralized network architecture in combination with distributed technology become more and more popular in recent years. Meanwhile, the blockchain with completely decentralizing features emerged and attract the attention of scholars.

The Blockstack[4] system can run on many famous blockchain platforms such as Bitcoin, and the system improves data storage performance by separating the data storage from the control layer. However, the Blockstack still rely on the existing IP network, which cannot meet the requirement of future network development. ODIN, A peer-to-peer open data index name system based on the blockchain, decentralized domain name management by using bitcoin sidechains. However, ODIN cannot be well adapted to massive data identifiers for large storage restrictions and slow reading due to directly storing data in the blockchain.

Although the solution mentioned above solve a few problems in the existing system, it still cannot meet the demand in Internet security due to the problems of low performance and application deployment. We believe that the next generation network should have features such as full-dimensional definition, diverse addressing routing. The topology, protocols, hardware, software, interfaces can be fully defined in the new network architecture [5].

identifier. Every node at domain boundary is responded for identifiers conversion and routing. the administration of Internet governance belongs to Internet participants from all over the world. They are no longer monopolized by an independent organization. The system is divided into three parts by a top-down hierarchical. The top-level domain of the network is made of top-level domain nodes from various countries government agencies keeping an alliance chain to achieve the consensus of the whole network. All resources in the network will be stored on the blockchain to ensure that the network resources are authentic and anti-tampering. Corresponding countries and

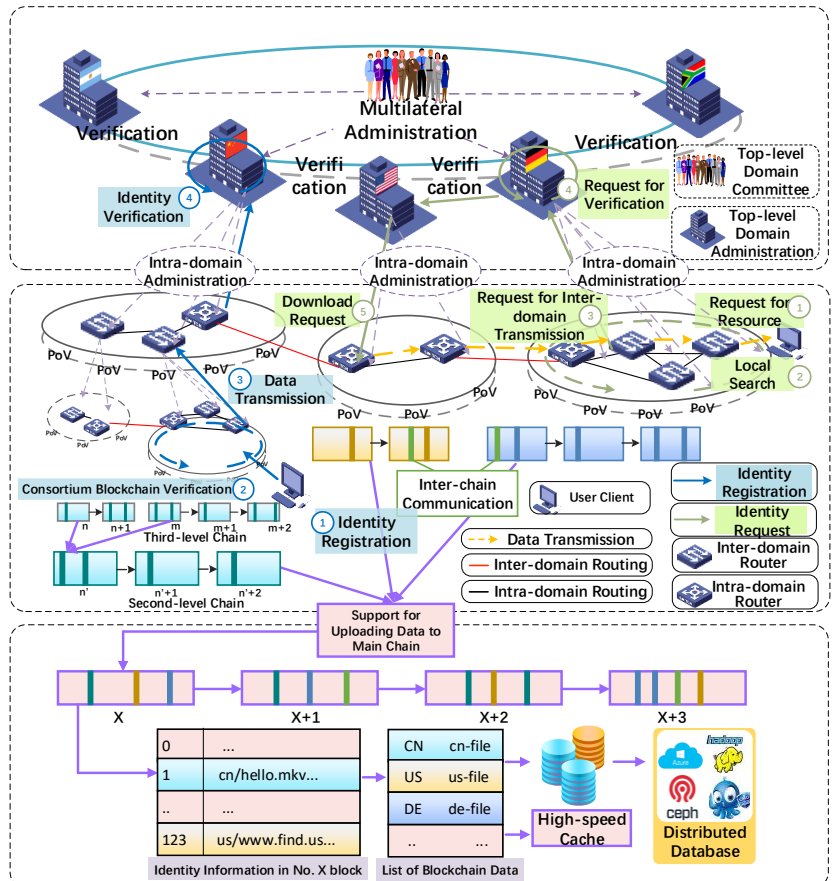

Figure 2. Multimodal network identifiers management architecture and operation flow

## III. OUR CONTRIBUTIONS

### A. System Architecture

The system proposed in this section establishes the multi-identifier network system realizing the characteristics of multilateral government [6][7][8]. Efficient transmission and operators' network deployment achieved in the prototype system proves the effectiveness of the system. The system makes a smooth transition from the existing network to new network architecture.

The framework of the multi-identifier network system with multilateral government system is shown in Fig. 1. Every node in the system maintain multi-identifier information including IP address identifier, content identifier, ID, geospatial information professional institutions manage the top-level domain and other domains for the identifiers management, identifiers registration. Consensus algorithm and identifiers management in the domain can be different from the other domain.

As shown in Fig. 2, there are network nodes such as supervisory nodes, individual users, and enterprise users. Network supervisory nodes are used as data access interfaces between the upper and lower domains to achieve hierarchical data transmission. Each domain has a corresponding supervisory node, which is mainly responsible for user management, identifier registration. Meanwhile, each node maintains multi-identifiers information table such as a content identifier, geospatial information identifier, ID and IP address identifier. The network layer in our system supports routing addressing where multiple identifiers such as ID, content, geospatial information, and IP address identifiers can be inter-translated in multi-identifier router. All the resources in the network have bound the identifier of the publisher. The resources publication and user registration information are recorded in the supervisory node of blockchain. The system, from the top to bottom, is divided into three layers, Control Layer, Routing Layer and Data Layer.

The upper control layer is responsible for domain name management and other transactions included offline affairs. After completing the block information verification and reaching the blockchain consensus, the routing status and the authentication in the domain will be recorded. It makes the whole network content of the system uniform, and has strong resistance and retrieving ability.

The middle layer of the proposed network is the routing layer is responsible for forwarding and filtering data packets. It also completes the operations of registering and parsing various network identifiers, such as content identifier, geospatial information identifier, ID and IP address identifier. Nodes at all levels rely on the voting consensus algorithm to complete the data consistency of each hierarchical Blockchain. The registering and resource transferring procedures for the identifier are shown in Fig. 2.

The efficient distributed relational database exists in the underlying data layer. The system has an efficient and distributed database. This includes the blockchain data sub-layer and the cloud storage sub-layer. The sub-layer of the blockchain data stores the minimum necessary data for routing, which is called on-chain data. The cloud storage sublayer stores all the information of the network identifier, which is called offline data, and the data is stored in a local database. Since the content and access behavior of all parties in the new network are effectively protected and managed, the behavior generated by their access to the network is undeniable. Any network attack or illegal behavior will also be recorded. Therefore, the use of these identifiers will make the network space in the orderly and secure state as well as guide the various traffic of the user to the new identifier bound with the identifier. Naturally, reduce IP network traffic. Information publishers will publish their data to the new identifiers, which will naturally guide the transformation of network traffic and systems to achieve IP-free.

### B. Key Technology

*1) An efficient and forkless consensus algorithm for the cooperative operation of the alliance*

Based on the framework of the theoretical basis of the network identifier management system, we introduce a new consensus algorithm called "Proof of Vote (PoV)" proposed by our team[9]. As shown in Fig. 3. The innovation lies in the separation of voting and billing. Without trusting a centralized institution, all the alliance vote together to conduct "decentralization" arbitration. Following the principle of "minority obeys the majority", the voting result is used as a legal proof that the system generates valid blocks by using the special identifier of the nodes in the alliance chain model. The idea of the voting proof is reflected in two designs in PoV consensus which is shown in Fig. 4.

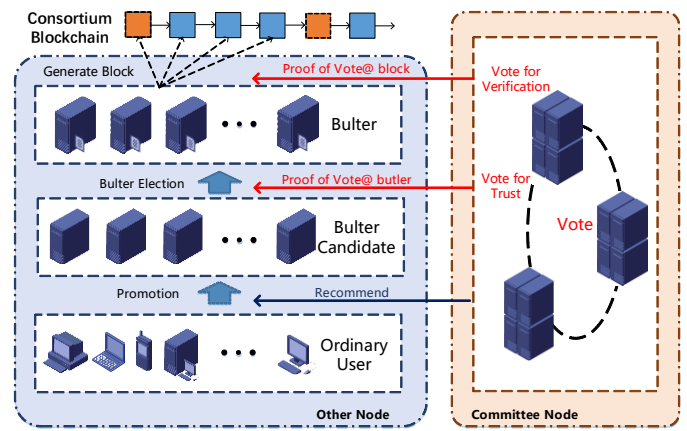

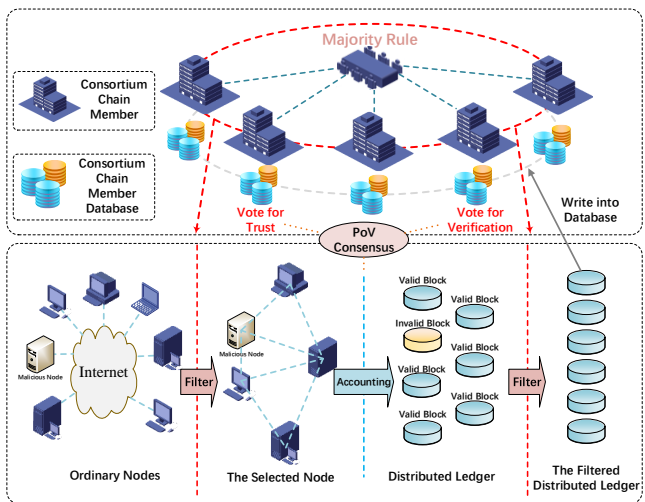

Figure 3. A consensus design for separating voting and accounting

Fig. 4. Concept of two kinds of voting in PoV

**Proof of Vote on Butlers**: For a vote of confidence in the butler, the total number of votes cast by the member for the butler at the end of each term indicates the confidence of the entire committee in the butler, and the reliability of each butler is demonstrated by the votes cast by the member for the butler.

**Proof of Vote on Blocks**: For a vote to verify the legitimacy of a block, more than half of the committee must verify each block in order to be considered a valid legal block. The result can be modified with the approval of more than half of the members if the system needs to modify a result. The legitimacy of each block can be proved by the voting result of the alliance.

Meanwhile, the PoV algorithm sets up the roles of butler and butler candidate. The stewarding team executes the decision results of the alliance chain. The butler team changes dynamically through decentralized voting, forming a voting consensus between the alliance nodes. The "majority voting result" can only confirm a final decision, making the alliance chain system run stable under the decentralized management of the alliance nodes.

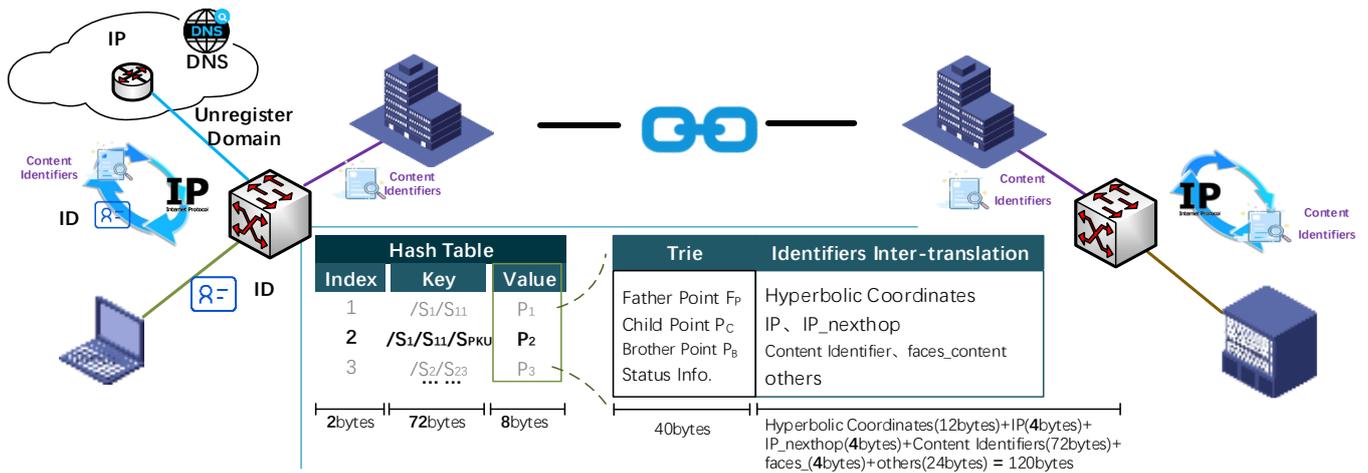

Fig. 5. The inter-translation scheme for multimodal identifiers

### 2) Multi-modal identifiers inter-translation scheme

In the future network, resources cannot be only limited to IP domain name identifiers. In our system, we introduce a multi-identifiers network space that integrates identifier, content, geospatial information and IP address. The resource may have one or more identifiers in this approach. Therefore, for identifier addressing and routing, it is particularly significate to design a set of multi-identifiers inter-translation scheme, which can translate the one identifier to another. The structure of identifiers inter-translate using Hash Prefix Table - Forwarding Information Base (HPT-FIB) is shown in Fig. 5. Where hash tables are used to support quick lookups and prefix trees are used to store the logical relationships between multi-identifiers.

Resource request steps are the following:
1. A user uses an identifier to request resource.
2. Multi-identifiers router response for resource query according to different identifier:
   (1) Traditional domain name: Direct use DNS query
   (2) IP address: If there is an identifier in the HPT-FIB, it shall be translated or forwarded; otherwise, the agent will access the traditional IP network
   (3) ID, content and other identifiers shall be firstly inquired in CS, PIT, and translation to publication. If they exist, they shall be translated or forwarded. Otherwise, they shall be forwarded to step 3.
3. If the identifier does not exist in the current domain, the multi-identifiers router will recursively query up to the top-level domain.
4. If there is no such identifier information in the top-level domain, it will be queried according to the subordinate domain specified by the identifier information until the lowest domain. The corresponding result will be returned if it exists. Otherwise, the query error message is returned.

In general, the length of the new network identifier is much longer than the IP address, so the multi-modal identifier addressing process will have more computing and storage costs, which will have a negative impact on the performance in a large-scale environment.

Table 1 Test for HPT-FIB entry generation

| Size of FIB table | 0.1 Billion | 1 billion | 2 billion | 2.5 billion | 3.0 billion | 3.5 billion |
|---|---|---|---|---|---|---|
| Run Time(s) | 187.58 | 1649.75 | 3723.98 | 4925.64 | 6271.49 | 7760.69 |
| Actual size of FIB table | 0.1 Billion | 1 billion | 2 billion | 2.5 billion | 3.0 billion | 3.5 billion |

The data source of the experiment for HPT-FIB is the self-generated ID, content and other identifiers. As it is a new type of network architecture, It is difficult to acquire large-scale data samples from real network traffic. Therefore, we analyze some statistical properties of the current network flow, using to obtain a large number of identifiers through simulation. Then we generate a large-scale FIB table on this basis realizing the evaluation and research of the scheme.

### 3) The new solution of network identity interworking and progressive deployment

In 2016, in order to promote the application of CCN [10]. It was proposed to use CCN as the tunnel technology to transmit IP packets [11]. The main innovation of this theory is to put forward the idea of "conversion agent", which is used for packet conversion between IP and CCN. However, the experiment still runs on IP, which is not a real sense of this tunnel technology. In 2017, Tsinghua University proposes the idea of "Dual-stack switch" to solve the progressive deployment CCN with IP[12]. However, this idea still stays the stage of theory, which is not verified by the real world experiment. In the same year, in order to make full use of the underlying resources, it was proposed for the first time that face in CCN should be dynamically mapped into MAC address [13]. It is considered to connect the content layer in CCN with the Ethernet link layer. Similarly, this theory is only in the conceptual stage and no real-world deployment has been made. In 2018, the concept of a connection in the content layer with the Ethernet link layer, directly achieve the transmission of data packets through the Ethernet link layer. Similarly, in 2018, the idea of using CCN as the tunnel technology to transmit IP packets through the Ethernet link layer was proposed. The key is that the interest packet was loaded with the IP packet. However, this idea is only under simulation stage through an IP network, not a real-world deployment.

Our system deploys CCN in the Ethernet link layer directly [14]. Not only our system is compatible with the IP network, but also can completely get rid of IP and run independently. The architecture of converged with Ethernet, TCP, and UDP is shown in Fig. 6. All the characteristics such as IP overlay CCN and multi-identifiers communication in our prototype system has been verified in a real-world network environment.

*4) Topology Approach of Achieving CAP Guarantee Bound for Consortium Blockchain Systems*

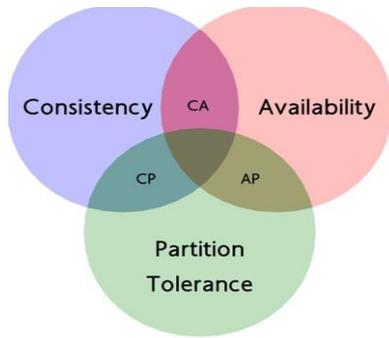

Fig. 9. Real-time monitoring on Blockchain node information

Mainstream Blockchain systems such as Bitcoin satisfy high availability and partition tolerance by weakening strict consistency to the final consistency within a period, which results in limited throughput performance of the system. Fig. 9 illustrates the triangular constraints between the three properties. Specifically, the following three performances cannot be strongly satisfied in a distributed system at the same time then, some high-performance Blockchain systems such as EOS choose to sacrifice part of the partition tolerance and gain the improvement of throughput performance by controlling the number of nodes participating in bookkeeping. These means only select two factors as the main points of reinforcement inconsistency, availability, and partition tolerance, while substantially weaken the third factor, leaving much room for the CAP guarantee bound.

Based on the consistency and availability of the consensus protocol used in the Blockchain-based system, we introduce the network topology designed to satisfy the strong partition tolerance in probability, to reach the limit of the coexistence of three factors in the CAP theorem. This topology construction method applies not only to the Blockchain system but also to some ''partially decentralized'' consortium block-chains. Our contributions are listed as follows:

1) We introduce the network topology structure into the design of Blockchain-based systems. By constructing a network topology suitable for the scale of the system with different requirements, it enhances partition tolerance to satisfy the coexistence of consistency, availability and partition tolerance as far as possible.

2) We propose an approach of calculating the partition tolerance probability and the average minimum repair time for the consensus protocol and network topology of the Blockchain-based system. The characteristics of the consensus protocol and the network topology are matched to the maximum extent by sampling and estimating each partition failure state.

3) We derive the partition tolerance of hierarchical network topology from the calculation approach of common network topology, which applies not only to the general single-chain architecture but also to multi-chain and cross-chain architectures.

## IV. PROTOTYPE AND EXPERIMENTS

### A. Description of Administrator Functions

The prototype system realizes the binding mechanism of user content and its private key signature and completes the development of the prototype system including the Blockchain administrator module and the client module. The management of Blockchain by administrator includes the real-time display of running state of nodes, the query of Blockchain data and operation and configuration of Blockchain nodes.

Management operation such as identifiers data, registered users are stored in each block of Blockchain, while specific user and identifier data are stored out of the chain. So the client needs to provide not only of blocks of data query function but also need to provide user data and identify the data query functions. For each registration and resource publishing of the client, it achieves consensus across the whole network, which is described as the number of transactions in the administrator interface and shown in Fig. 10.

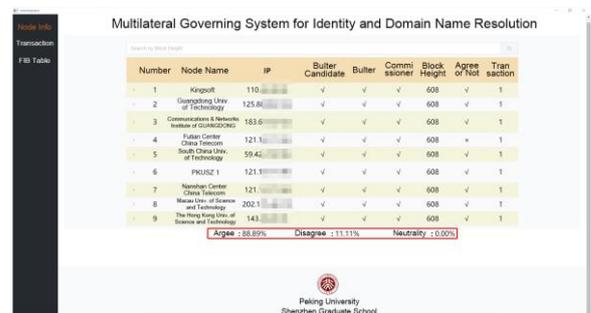

Fig. 10. Real-time monitoring on Blockchain node information

Multi-identifier routing nodes take the responsibility of transformation and routing navigation between different identifiers. Each node maintains an internal FIB table, routing table, and other routing information. While the interface realizes the function of multi-identifier routing node status monitoring and multi-identifier routing node data display. The node routing status information is shown in Fig. 11.

Fig. 11. Node routing status information

*B. Description of Client Functions*

As shown in Fig. 12. The client of the user's web page offers users the processes from registration and publishing resources to resources querying and obtaining. Currently, only the video data is available in the network resources. User client provides the function of video online streaming.

To achieve the objectives described above, the web functions of the user's web client include:

1. User registration: Allows the user to register a unique prefix and generate a key for the user to log in to the system to publish resources. User registration is bound to the user's real identifier information.

2. Resource publishing: Registered users can publish a network identifier under their registered prefix with the true location of the network identifier. When a user publishes a resource, he provides the key and binds the hash value of the published resource to ensure the non-repudiation of the published resource.

3. Resource query: a webpage is needed to provide users with resource navigation and retrieval

Fig. 12. User registration and resource publishing interface

*C. The experiment of Multi-modal Network Identifier Access*

This system takes the lead in implementing CCN in the MAC layer, which is independent of the IP network environment but compatible with existing IP network. The prototype system has completed the identifier generation, management and resolution functions of the multi-modal system as well as the identifiers exchange functions under different network scenarios, including IP-CCN-IP, IP-CCN, CCN-IP, CCN-IP-CCN, and CCN-CCN, realizing the conversion of address identifiers and content identifiers. The prototype system supports the transmission of real-world video streams under multiple transmission channels providing by several ISP.

The topological relationship of the prototype is shown in Fig. 13. It is consist of Peking University Shenzhen Graduate School, China Telecom, China Unicom, Guangdong Institute of new generation communication and network innovation, Kingsoft Corp., South China University of Technology, Guangdong University of Technology, the Hong Kong University of Science and Technology, the Chinese University of Hong Kong, Macau University of Science and Technology. The prototype system is deployed on the real Internet for testing function of multi-identifier registration and resolution.

*1) Performance eveluation on IP-CCN-IP scenario*

The resources were uploaded on pkusz1 node of China Telecom Futian data center and pkusz3 node of China Telecom Nanshan data center. Test of video streaming is conducted through node9 agent of Peking University Shenzhen Graduate School.

The video data on pkusz1 in Futian data center is pulled from node9 through the agent. The experimental topological relation and transmission result was shown in Fig. 14, and the rate is maintained at 2.65MB/s (21.2Mbps).

The video data on pkusz1 in Nanshan data center is pulled from node9 through an agent. The experimental topological relation and transmission result was shown in Fig. 15, and the rate is maintained at 2.44MB/s (19.52Mbps).

We pulled the video data from Futian pkusz1 and Nanshan data center pkusz3 simultaneously through agent node9 and tested the topological relationship and the transmission results are shown in Fig. 16 with rates of 1.18MB/s (9.44 Mbps) and 1.71MB/s (13.68 Mbps) respectively.

Compared with pulling resources alone, the rate decreases, Compared with pulling resources separately. but the total transmission rate remains the same. The main reason is that the bandwidth of node9 node is limited, which becomes a performance bottleneck.

In order to reflect the feasibility of identifiers transmission in different network environments provided by different ISP, we put the data on the pkusz3 node in Nanshan data center and conduct the experiment of video streaming transmission through the Server2 agent node of Macau University of Science and Technology. Affected by the export bandwidth of Macau University of Science and Technology. The average speed is 1.2MB/s (9.6 Mbps), but it can still stream HD video. The experimental topological relationship and test results are shown in Fig. 17.

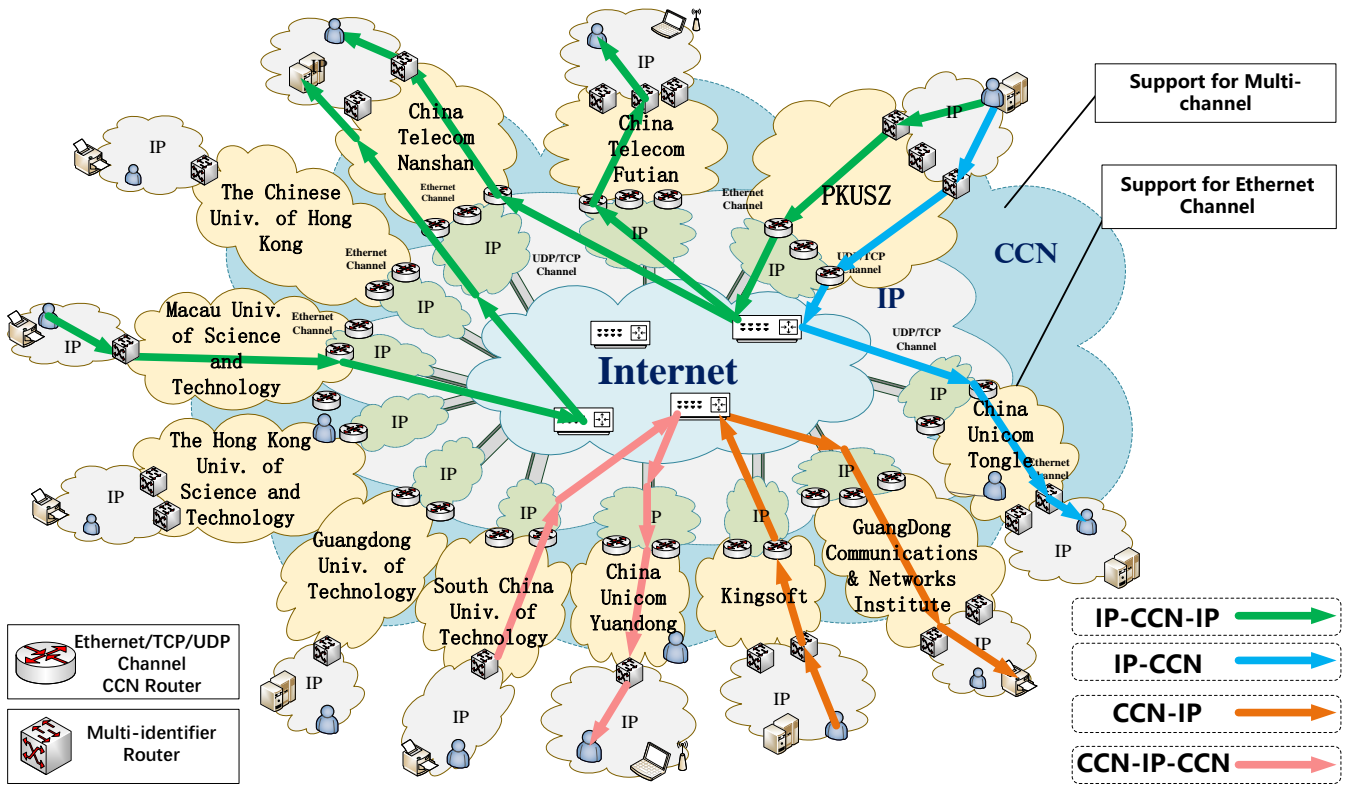

Figure 13. Overview of prototype testing system

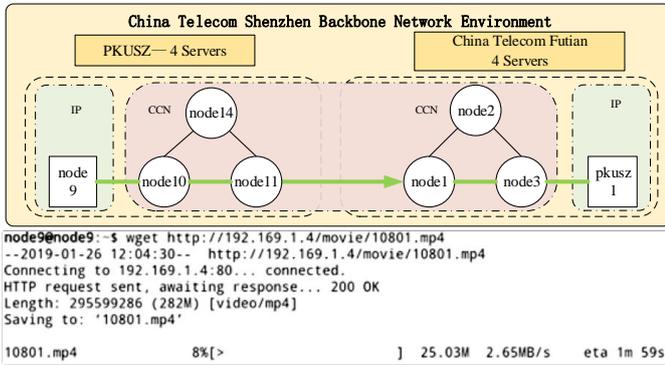

Fig. 14. Overview of prototype testing scenario

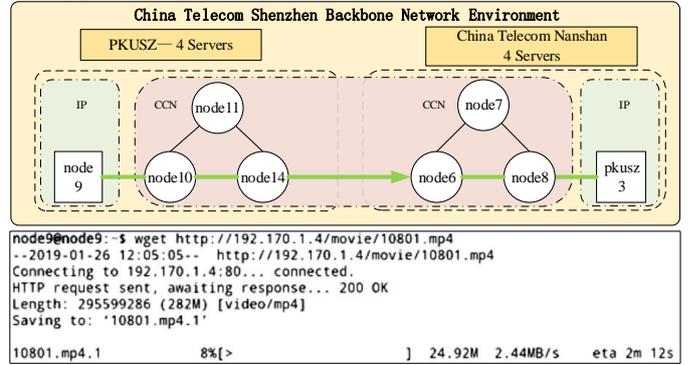

Fig. 15. Overview of prototype testing scenario

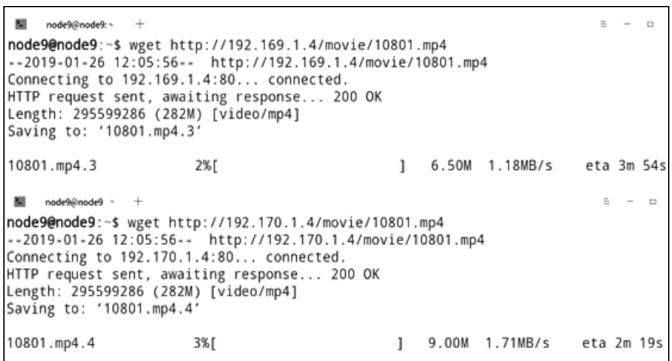

Fig. 16. Transmission speed of prototype testing scenario

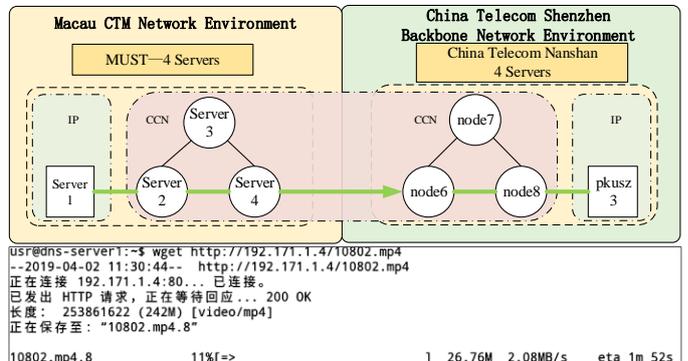

Fig. 17. Overview of prototype testing scenario

## 2) Performance evaluation on IP-CCN scenario

The resource is upload on host2 node of China Unicom Tongle data center. Node10 agent of Peking University Shenzhen Graduate School is used as a multi-identifier router. Test of video streaming is conducted through different network environment provided by two ISPs. The local host runs on the IP network. While host2 runs on CCN network. The average speed is 1.2MB/s (9.6 Mbps). The experimental topological relationship and test results are shown in Fig. 18.

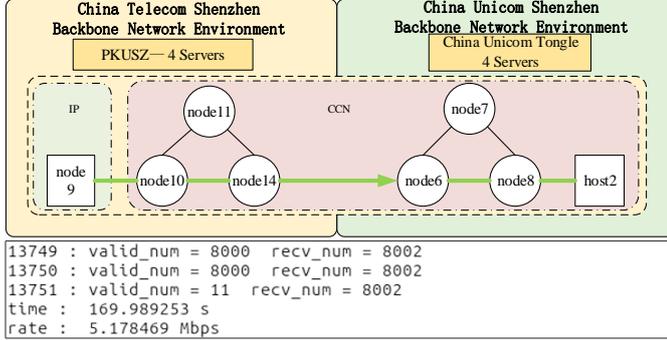

Fig. 18. Overview of prototype testing scenario

## 3) Performance evaluation on CCN-IP scenario

The resource is placed on the gdcni1 of GDCNI in IP environment. We take node10 of Peking University Shenzhen Graduate School as the multi-identifier router. Kingsoft host1 in CCN network environment pulls the resources on node9. The experimental topological relationship and test results are shown in Fig. 19. The average rate of the video stream is 1MB/s (8Mbps).

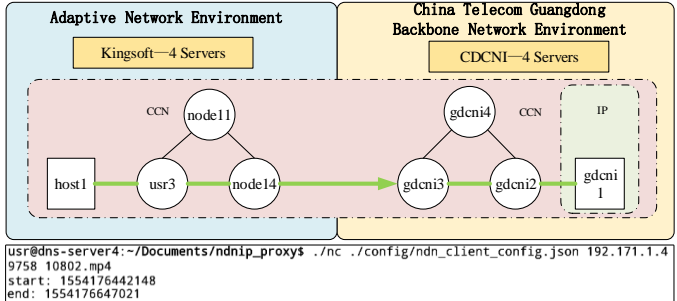

Fig. 19. Overview of prototype testing scenario

## 4) Performance evaluation on CCN-IP-CCN scenario

The resource is placed on host2 node of China Unicom Yuandong in CCN environment. We use node10 of Peking University Shenzhen Graduate School in CCN network environment to pull resources on Kingsoft host1. The experimental topological relationship and test results are shown in Fig. 20. The average rate of the video stream is 2.65MB/s (21.2Mbps).

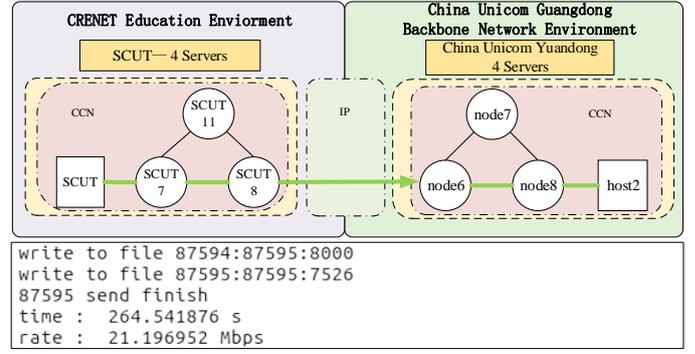

Fig. 20. Overview of prototype testing scenario

## V. CONCLUSION

With the implementation of full-dimensional definability under the open network architecture, the functions and business mechanisms of the network are no longer limited to the existing architecture and protocols, greatly improving the network service capability. Based on the considerations above, we design a new domain name management system with the characteristics of multilateral co-management, multi-modal addressing, efficient availability, and privacy protection. The system's key technologies have obtained nearly 20 PCT patents and 7 US authorized patents. Our prototype consists of Peking University Shenzhen Graduate School, China Telecom, China Unicom, Guangdong Institute of new generation communication and network innovation, Kingsoft Corp., South China University of Technology, Guangdong University of Technology, the Hong Kong University of Science and Technology, the Chinese University of Hong Kong, Macau University of Science and Technology. The prototype system is deployed on the real Internet testing function of multi-identifier registration and resolution. The deployment and test results show that the system scheme is completely feasible. In addition, the results also show that the system performance is excellent by video stream transmission pressure test.

## VI. ACKNOWLEDGMENTS

The design and development of the Multilateral Governing System for identifier and Domain Name Resolution in Multi-modal Internet are strongly supported by the participating organizations. At the same time, I would like to thank the participated specialist for their contribution to prototype system testing in the past six months.

The system is supported by the following program:
1. National Key R&D Program of China (2016YFB0800101 and 2017YFB0803204)
2. National Natural Science Foundation of China (No. 61671001 and 61521003)
3. Guangdong Key Program GD2016B030305005
4. Shenzhen Research Programs (JSGG20150331101736052, ZDSYS201603311739428, JCYJ20170306092030521).

5. Shenzhen Municipal Development and Reform Commission (Disciplinary Development Program for Data Science and Intelligent Computing)

## REFERENCES

[1] Lan J L, Cheng D N, Hu Y X. Research on reconfigurable information communication basal network architecture. Journal on Communications, 2014, 35(1): 128-139.
[2] Wu J X. Thoughts on the development of novel network technology (in Chinese). Sci Sin Inform, 2018, 48: 1102-1111.
[3] Wu J P, Lin S, Xu K, et al. Advances in evolvable new generation internet architecture. Chinese journal of computers, 2012, 35(6):1094.
[4] ALI M, NELSON J, SHEA R, et al. Blockstack: a global naming and storage system secured by block chains. USENIX Annual Technical Conference. 2016: 181-194
[5] Lv P, Liu Q R, Wu J X, et al. New generation software-defined architecture (in Chinese). Sci Sin Inform, 2018, 48:315-328
[6] Li H, Wang X G, Lin Z L, et al. US Patent, 10178069, 2019-01-08
[7] Li H, Li K J, Chen Y L, et al. US Patent, 15/997,710, 2019-03-14
[8] Li H, Wu J X, Zhang X C, et al. PCT No.CN2019/073507, 2019-01-28
[9] Li K J, Li H, Hou H X, et al. Proof of Vote: A High-Performance Consensus Protocol Based on Vote Mechanism & Consortium Blockchain. In: Proceedings of IEEE Proof of Vote: A High-Performance Consensus Protocol Based on vote Mechanism & Consortium Blockchain. 2017.
[10] Jacobson V, Smetters D K, Thornton J D, et al. Networking Named Content. Communications of the ACM, 2012,55(1):117-124.
[11] Moiseenko I. and Oran D. TCP/ICN: Carrying TCP over Content Centric and Named Data Networks[C]//Proceedings of the 3rd ACM ICN, 2016: 112-121.
[12] Wu H, Shi J X, Wang Y X, et al, On Incremental Deployment of Named Data Networking in Local Area Networks[C]//Proceedings of the ACM/IEEE ANCS, 2017: 82-94.
[13] Kietzmann P, Gündoğan C and et al. The Need for a Name to MAC Address Mapping in NDN: Towards Quantifying the Resource Gain[C]//Proceedings of the 4th ACM ICN, 2017: 36-42.
[14] Li H, Xu R, Wu J X, et al. PCT No.CN2018/120686, 2018-12-12